\documentclass[aps,prb,twocolumn,preprintnumbers,amsmath,amssymb,superscriptaddress]{revtex4-1}
\usepackage{graphicx}
\usepackage{subfigure}
\usepackage{epsfig}
\usepackage{dcolumn}
\usepackage{bm}
\usepackage{ulem}
\usepackage{color}

\def\avg#1{\left\langle#1\right\rangle}
\def\bra#1{\left\langle#1\right|}
\def\ket#1{\left|#1\right\rangle}

\def\be{\begin{equation}}       \def\ee{\end{equation}}
\def\bea{\begin{eqnarray}}      \def\eea{\end{eqnarray}}
\def\ba{\begin{array} }
\def\ea{\end{array} }
\def\bnum{\begin{enumerate} }
\def\enum{\end{enumerate}}

\def\nn{\nonumber}

\def\=>{\Rightarrow}
\def\>{\rightarrow}

\def\eye2{Fathbb{I}}

\def\Eq#1{Eq.~(\ref{#1})}
\def\Fig#1{Fig.~\ref{#1}}

\def\Tr{\mathrm{Tr}}

\renewcommand{\>}{\rangle}

\begin{document}

\title{Fermion-sign-free Majorana-quantum-Monte-Carlo studies of quantum critical phenomena of Dirac fermions in two dimensions}
\author{Zi-Xiang Li}
\affiliation{Institute for Advanced Study, Tsinghua University, Beijing, 100084, China}
\author{Yi-Fan Jiang}
\affiliation{Institute for Advanced Study, Tsinghua University, Beijing, 100084, China}
\affiliation{Department of Physics, Stanford University, Stanford, CA 94305, USA}
\author{Hong Yao}
\email{yaohong@tsinghua.edu.cn}
\affiliation{Institute for Advanced Study, Tsinghua University, Beijing, 100084, China}
\affiliation{Collaborative Innovation Center of Quantum Matter, Beijing 100084, China}

\begin{abstract}
Quantum critical phenomena may be qualitatively different when massless Dirac fermions are present at criticality. Using our recently-discovered fermion-sign-free Majorana quantum Monte Carlo (MQMC) method introduced by us in Ref. [\onlinecite{ZXLi-14}], we investigate the quantum critical phenomena of {\it spinless} Dirac fermions at their charge-density-wave (CDW) phase transitions on the honeycomb lattice having $N_s=2L^2$ sites with largest $L=24$. By finite-size scaling, we accurately obtain critical exponents of this so-called Gross-Neveu chiral-Ising universality class of {\it two} (two-component) Dirac fermions in 2+1D: $\eta=0.45(2)$, $\nu=0.77(3)$, and $\beta=0.60(3)$, which are qualitatively different from the mean-field results but are reasonably close to the ones obtained from renormalization group calculations.
\end{abstract}
\date{\today}

\maketitle
\section{Introduction}
Quantum phase transitions have attracted enduring interests in condensed matter for many decades[\onlinecite{Sachdev-book}]. For contiguous phase transitions, scaling invariance and universality classes were introduced as fruitful concepts to understand them[\onlinecite{Kadanoff-chapter}]. Renormalization group (RG) provides us a modern and powerful framework to characterize critical behavior and university class[\onlinecite{Fradkin-book},\onlinecite{Cardy-book}]. When gapless fermions are present and coupled with order parameter at criticality, quantum critical phenomena may be dramatically changed; characterizing them are often challenging due to strong couplings between gapless fermions and order parameter. Such fermionic quantum critical points include quantum spin-density-wave/charge-density-wave/nematic transitions in metals with Fermi surface or in semimetals with massless Dirac fermions.

Interaction-induced quantum phase transitions in two-spatial-dimension semimetals with $N$ {\it two-component} massless Dirac fermions are particularly interesting in many systems such as graphene[\onlinecite{Castro Neto-10}], surfaces of topological insulators[\onlinecite{Hasan-2010},\onlinecite{Qi-2011}], and d-wave nodal superconductors[\onlinecite{Sachdev-book}]. Usually, massless Dirac fermions are gapped by breaking certain symmetry such as inversion when interactions are larger than a critical value. Such quantum critical behaviors at low energy may be described by the Gross-Neveu[\onlinecite{Gross-Neveu-74}] or Gross-Neveu-Yukawa theory which has been studied in graphene[\onlinecite{Herbut-06,Herbut-09a,Herbut-09b,Herbut-14}], in d-wave nodal superconductors[\onlinecite{Laughlin-98,Vojta-98,Franz-02,Kim-08}], as well as in high-energy physics[\onlinecite{Rosenstein-93,Karkkainen-94,Rosa-01,Hofling-02}] using various RG approaches such as large-$N$ and dimensional regularization. Numerical methods such as quantum Monte Carlo simulations[\onlinecite{Scalapino-81,Blankenbecler-81,Rebbi-81,Hirsch-81,Hirsch-85}] have also been employ to study this type of quantum critical behavior in lattice models[\onlinecite{Sorella-92,Sorella-12,Chandrasekharan-13, Assaad-13, LWang-14a,Toldin-14}]. Nonetheless, critical exponents obtained in RG calculations are often in discrepancy with the ones from QMC calculations of lattice models when QMC simulations encounter the notorious fermion-sign-problem[\onlinecite{Loh-90},\onlinecite{Troyer-05}] or when $N$ is small.

The smallest number of two-component Dirac fermions realizable in 2d lattice models is $N=2$, a prototype example of which is from spinless fermions on the honeycomb lattice (note that $N=1$ massless Dirac fermions in 2+1D can only be realized on surfaces of 3d topological insulators). In the presence of strong repulsion between fermions on nearest-neighbor (NN) sites, the ground state develops a finite charge-density-wave order. Close to transition, chiral Dirac fermions strongly couple with Ising order parameter, rendering this chiral-Ising transition in 2+1D qualitatively different from the usual Ising transition in 2+1D. Recently, a remarkable continuous-time (CT) QMC study[\onlinecite{LWang-14a}] of the CDW transition in the honeycomb model of $N_s=2L^2$ sites with $L$ up to 15 obtained $\eta=0.30(2)$ which is in some discrepancy with previous RG results ($\eta\approx 0.50\sim 0.64$)[\onlinecite{Rosenstein-93,Karkkainen-94,Rosa-01,Hofling-02}].

We recently discovered an auxiliary-field QMC method employing Majorana representation which can solve the fermion-sign problem in such spinless fermion models[\onlinecite{ZXLi-14}]. Using this sign-free Majorana quantum Monte Carlo (MQMC), we performed highly-accurate simulations of spinless fermions on the honeycomb lattice of $N_s=2L^2$ sites with largest $L=24$ and obtained critical exponents of the $N=2$ chiral-Ising transition in 2+1D: $\eta=0.45(2)$, $\nu=0.77(2)$, and $\beta=0.60(3)$. Note that we would obtain a much smaller $\eta=0.32(2)$ which is consistent from previous CT-QMC simulations[\onlinecite{LWang-14a}], if we were using simulation results on lattices only up to $L=15$. Using MQMC, we also investigated the CDW transition in the $\pi$-flux square lattice model with NN repulsions on lattices of $2L^2$ sites with largest $L=24$ and obtained similar critical exponents. Our results are in reasonable agreement with the existing RG calculations of the Gross-Neveu-Yukawa theory: $\eta\approx 0.50\sim 0.64$ and $\nu\approx 0.74\sim 0.93$ [\onlinecite{Herbut-14},\onlinecite{Rosenstein-93},\onlinecite{Rosa-01},\onlinecite{Hofling-02}] even though there are still slight discrepancies. Our numerically-exact results may serve as a possible benchmark for higher-order RG calculations or other theoretical analysis in the future.

\section{Models}
To investigate the universal quantum critical behavior of the $N=2$ chiral-Ising transition in 2+1D, we study the following simple interacting spinless fermion model on the honeycomb lattice:
\bea
H&=& -\sum_{\avg{ij}}\left[t_{ij} c^\dagger_i c_j + h.c.\right]+\sum_{\avg{ij}} V (n_i - \frac12)(n_j - \frac12), ~~~\label{ham1}
\eea
where $c^\dag_i$ creates a fermion on site $i$, $t_{ij}=t$ is the hopping of fermions between NN sites, and $V$ is the density interaction between fermions on NN sites. Here the hopping and interaction parts are labeled by $H_0$ and $H_\textrm{int}$, respectively. Hereafter we set $t=1$ as the unit of energy.

Noninteracting fermions on the honeycomb lattice features massless Dirac dispersion around two inequivalent Dirac points $\pm \vec K$. Because of vanishing density of states, such $N=2$ two-component mass Dirac fermions are stable against any weak interactions which preserving symmetries of the model but can be gapped when interactions are sufficiently strong. In this paper, we shall focus on the CDW transition induced by relatively-strong NN repulsions. Note that quantum anomalous Hall (QAH) ordering will be induced by strong next-nearest-neighbor (NNN) repulsions[\onlinecite{Raghu-08}] and pair-density-wave ordering will be induced by relatively-strong NN attractions [\onlinecite{SKJian-14}], both of which are not the focus of present numerical studies.

\section{The MQMC method}

As an intrinsically-unbiased numerical method, QMC[\onlinecite{Scalapino-81, Blankenbecler-81, Rebbi-81, Hirsch-81, Hirsch-85}] is an important and valuable tool to study quantum phases of matter and critical behaviors at phase transitions when it is free from the fermion-sign problem. To access the quantum critical behavior of the CDW transition induced by repulsions, it is desired to simulate the interacting model using fermion-sign-free QMC on lattices with relatively-large size. Nonetheless, conventional auxiliary-field QMC simulations of interacting spinless fermion model generally suffer from the fermion-sign problem because the usual strategies[\onlinecite{Hirsch-86,Hands-00,Wu-03,Wu-04,Wu-05,Assaad-05,Meng-10,Berg-12,Assaad-12,Wu-12}] employed in auxiliary-field QMC for solving fermion-sign problem of even species of fermions do not directly work. Recently, we discovered that certain interacting spinless fermion systems can be studied by fermion-sign-free QMC employing Majorana representation[\onlinecite{ZXLi-14}]. Upon observing that each complex fermion can be represented as two Majorana fermions, we perform Hubbard-Stratonovich (HS) transformations to decouple interactions in terms of bilinear Majorana fermions. Under certain conditions such as particle-hole symmetry, a symmetric treatment of two species of Majorana fermions is possible such that the Boltzmann weight is a product of two identical real quantities and is then positive definite.

To the best of our knowledge, MQMC is the first QMC approach based on auxiliary fields to solve fermion sign problem in a class of interacting spinless fermion models[\onlinecite{ZXLi-14}]. We emphasize that the auxiliary-field MQMC approach proposed here is qualitatively different from the sign-free continuum-time QMC method[\onlinecite{Wiese-95,Wiese-99,Shailesh-13,Shailesh-14,Shailesh-14,Iazzi-14}]. One important advantage of the fermion-sign-free MQMC is its high-efficiency of directly simulating interacting models at zero temperature by using projector. To be self-contained, here we briefly review the fermion-sign-free MQMC algorithm introduced by us in Ref.[\onlinecite{ZXLi-14}]. The partition function after Trotter decomposition is given by $Z = \Tr\left[e^{-\beta H}\right]\simeq\Tr\left[\prod_{n=1}^{N_\tau} e^{-H_{0}(n)\Delta\tau}e^{-H_{\textrm{int}}(n)\Delta\tau}\right]$,
where $n$ labels the discrete imaginary time, $\Delta\tau N_\tau= \beta$, and the approximation is good for small enough $\Delta\tau$. We observe that it will be advantageous to write complex fermion operators in Majorana representation: $c_{i}=\frac{1}{2}(\gamma_{i}^{1}+i\gamma_{i}^{2}), ~c^{\dagger}_{i}=\frac{1}{2}(\gamma_{i}^{1}-i\gamma_{i}^{2})$,
which enable us to rewrite \Eq{ham1} as follows:
\bea\label{ham2}
H= \sum_{\avg{ij}} \frac{it}{2}(\gamma^{1}_{i}\gamma^{1}_{j} + \gamma^{2}_{i}\gamma^{2}_{j})
- \frac{V}{4}\sum_{\avg{ij}}(i \gamma_{i}^{1}\gamma_{j}^{1})(i \gamma_{i}^{2}\gamma_{j}^{2}),
\eea
where gauge transformations $c_{i}\rightarrow i c_{i}$ for $i\in $ A sublattice were implicitly made so that $H_0$ can be written symmetrically in the two components of Majorana fermions.

Now, it is clear that we should perform HS transformations to decouple $H_\textrm{int}$ in following way:
\bea
&& e^{\frac{V\Delta\tau}{4}(i \gamma_{i}^{1}\gamma_{j}^{1})(i\gamma_{i}^{2}\gamma_{j}^{2})}
= \frac{1}{2}\sum_{\sigma_{ij}=\pm 1}e^{\frac12\lambda \sigma_{ij} (i \gamma_{i}^{1} \gamma_{j}^{1} + i \gamma_{i}^{2}\gamma_{j}^{2})-\frac{V\Delta\tau}{4}}, \label{V1sign}~~~
\eea
where $\lambda$ is a constant with $\cosh \lambda = e^{\frac{V\Delta\tau}{2}}$. The free-fermion Hamiltonian after the HS transformations is a sum of two parts each of which involves only one component of Majorana fermions. This makes the solution of fermion-sign problem possible in the MQMC. The partition function is a sum of Boltzmann weight which depend on auxiliary field configurations $\{\sigma\}$ in space-time: $Z=\sum_{\{\sigma\}} W(\{\sigma\})= \prod_{a=1}^2 W_a(\{\sigma\})$ with
\bea\label{trace0}
W_a(\{\sigma\})&=&\Tr\left[ \prod_{n=1}^{N_\tau} e^{\widetilde \gamma^a h^a (n) \gamma^a} \right],
\eea
where $\widetilde \gamma^a$ represents the transpose of $\gamma^a$ and $h^a(n)$ is a $N_s\times N_s$ matrix ($N_s=$ the number of lattice sites): $
h^a_{ij}(n) =i\left[t\Delta\tau \delta_{\avg{ij}}+\lambda\sigma_{ij}(n)\delta_{\avg{ij}} \right]$
with $\delta_{\avg{ij}}=\pm 1$ for NN $ij$ sites and 0 otherwise. The Boltzmann weight can be proven to be always positive. We note that $\hat h^1(n)\equiv \widetilde \gamma^1 h^1(n)\gamma^1$ of Majorana fermions $\gamma^1$ can be identically mapped to $\hat h^2(n)\equiv \widetilde \gamma^2 h^2(n)\gamma^2$ by the time-reversal transformation $T: \gamma_i^1 \to (-1)^i \gamma_i^2$.
Consequently, we obtain
\bea
W_{1}(\{\sigma\})= W_2^{*}(\{\sigma\}),
\eea
which renders the Boltzmann weight $W(\{\sigma\}) = W_{1}(\{\sigma\}) W_2(\{\sigma\})\ge 0$ for any auxiliary field  configuration $\{\sigma\}$. Explicitly, we obtain the positive-definite Boltzmann weight  $W(\{\sigma\})=\big |\det\big[ \mathbb{I}+\prod_{n=1}^{N_\tau} e^{h^a(n)}
\big]\big|$,
where $a=1$ or $2$ giving rise to the same result.

Because we are interested in the quantum phase transition at zero temperature, we can use projector[\onlinecite{Sugiyama-86,Sorella-89,White-89}] MQMC to explore the ground state properties of the system. In the projector QMC, a trial wave function $\ket{\psi_T}$ is introduced and after projection the ground state $\ket{\psi_0}$ can be obtained. Consequently, the expectation values of observable $\hat O$ is:
\bea
\frac{\bra{\psi_{0}} \hat O \ket{\psi_{0}}}{\bra{\psi_{0}}\psi_{0}\rangle} = \lim_{\theta \rightarrow \infty } \frac{\bra{\psi_{T}} e^{-\theta H} \hat O e^{-\theta H} \ket{ \psi_{T}}}{\bra{\psi_{T}} e^{-2 \theta H}  \ket{\psi_{T}}},
\eea
In our simulations, we use an Slater-determinant wave function as the trial wave function $\psi_{T}$. Similarly, we can prove that projector MQMC is also free from fermion-sign-problem in the parameter region where $V>0$, as shown in details in Ref.[\onlinecite{ZXLi-14}].

\section{Numerical MQMC Results}
\subsection{The CDW transition on the honeycomb lattice}
\begin{figure}[b]						
\centering
\subfigure{\includegraphics[width=4.25cm]{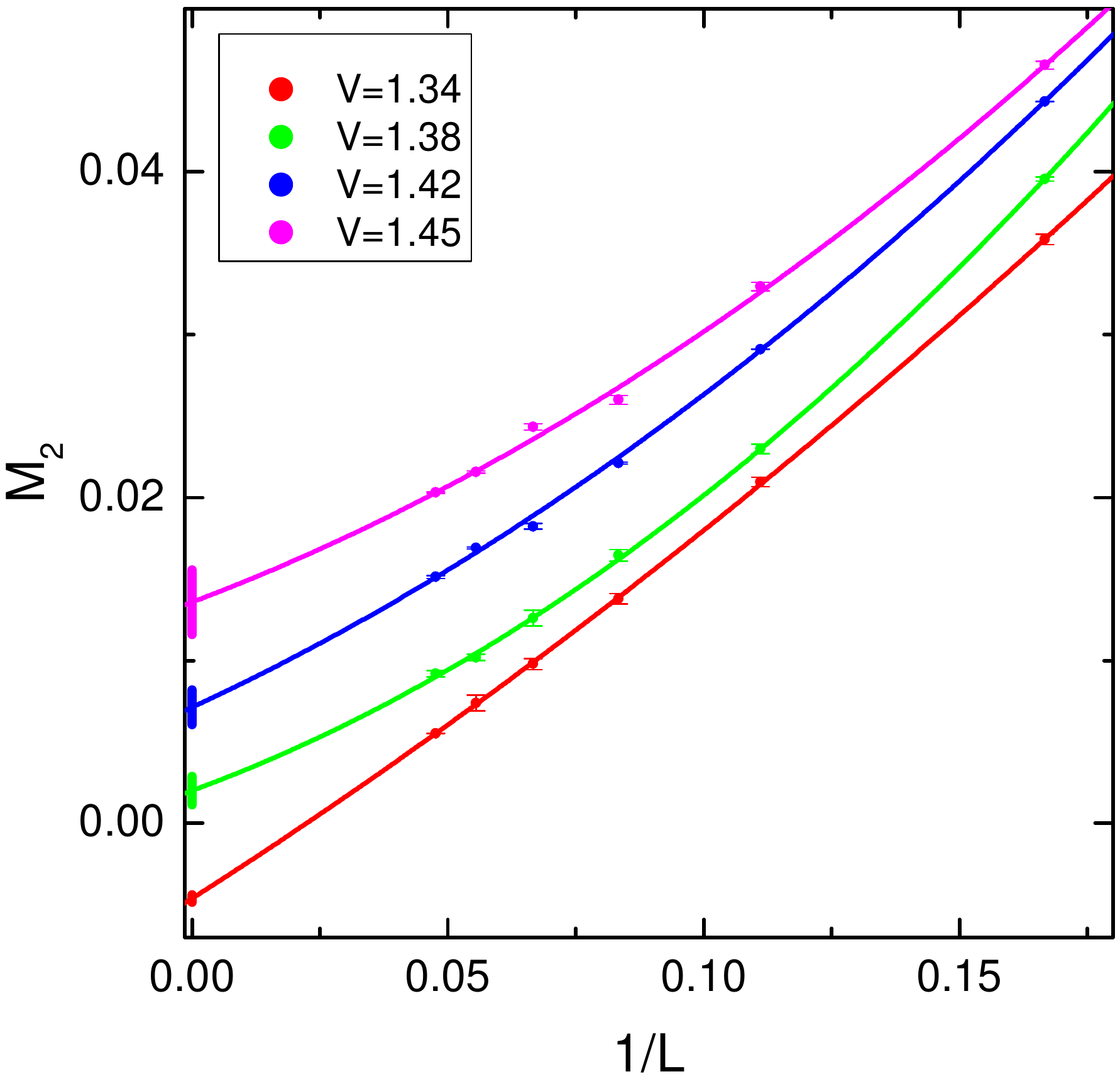}}
\subfigure{\includegraphics[width=4.25cm]{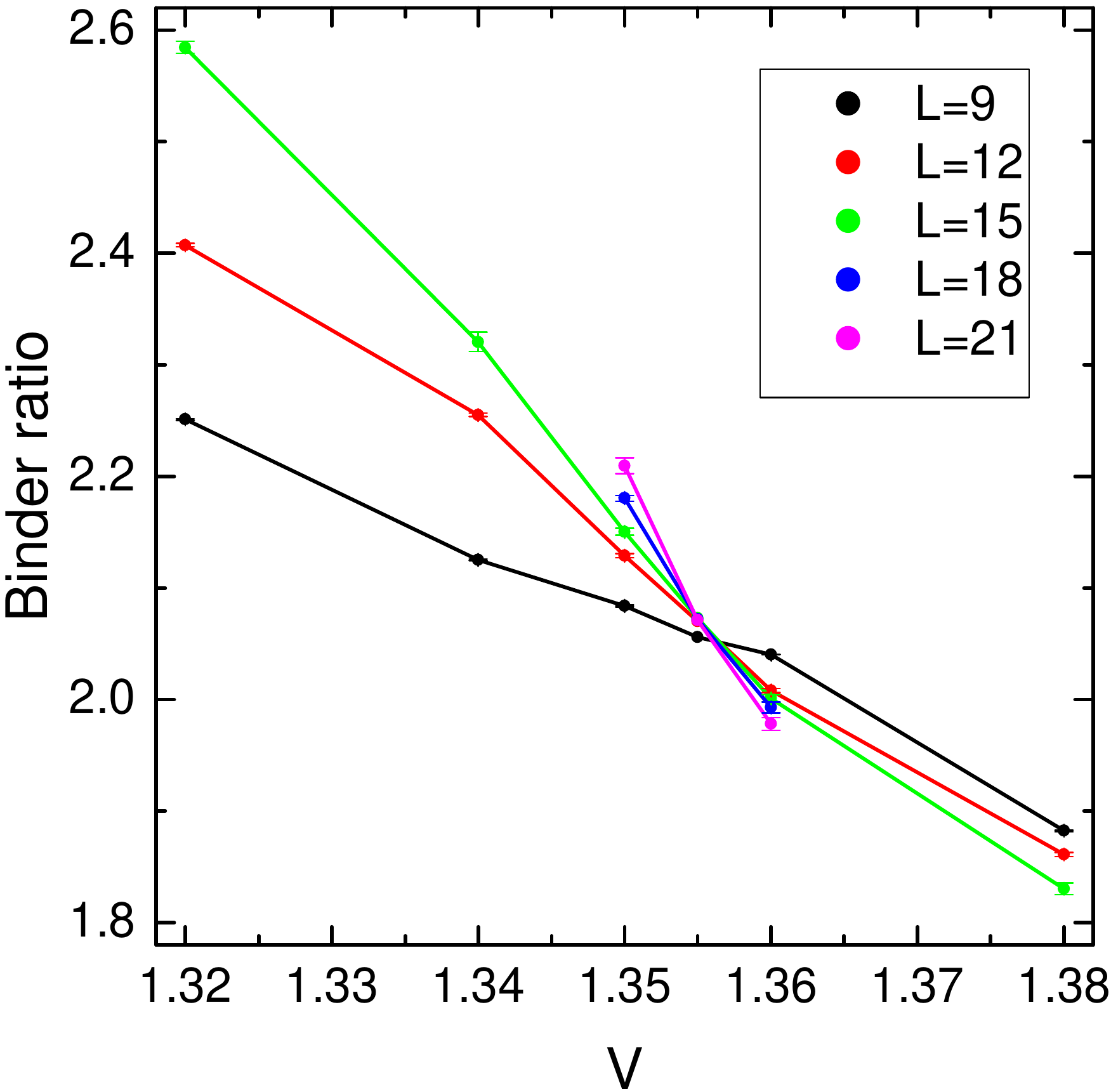}}
\caption{(a) For the honeycomb model, finite-size scaling of the CDW structure factor $M_2$ obtained in the projector (zero-temperature) MQMC simulations on lattice of $N_s=2L^2$ sites for various $V$ and $L=6\sim 21$. (b) The Binder ratios $B\equiv M_4/M_2^2$ for various interaction $V$ and various $L=9\sim 21$, are plotted. The crossing of Binder ratios shows that the critical interaction for the CDW transition is $V_{c}=1.355(1)$.  }
\label{honeycombscaling}
\end{figure}
We have performed fermion-sign-free MQMC simulations to study the spinless fermion model on the honeycomb lattice with NN repulsive interaction $V$. When $V>V_c$, the ground state develops a finite CDW ordering which breaks the inversion symmetry of the model. To find the critical interaction $V_c$ of the CDW phase transition, we compute the CDW structure factor on a finite lattice using MQMC:
\bea
M_2 = \sum_{ij} \frac{\eta_i \eta_j}{N_s^{2}} \big\langle(n_i - \frac{1}{2})(n_j - \frac{1}{2})\big\rangle,
\eea
where $N_s=2L^2$ is the total number of lattice sites and $\eta_i = \pm 1$ for $i\in  A(B)$ sublattice. The CDW order parameter $\Delta_\textrm{CDW}$ can be obtained through $\Delta_\textrm{CDW}^2=\lim_{L\to \infty} M_2$. We perform finite-size scaling analysis for $M_2$ by fitting $M_{2}$ to a second-order polynomial in $1/L$, namely $a_0 + \frac{a_1}{L} + \frac{a_2}{L^2}$, and then extrapolate to the thermodynamics ($L\to\infty$) limit to identity whether the system is the semimetal or CDW phase. The results are shown in \Fig{honeycombscaling}(a) for various $V$. The extrapolation of $M_{2}$ to $L=\infty$ shows that the critical value $V_c$ of the CDW transition should be between 1.34 and 1.38. Note that the critical point $V_c$ determined in this way is normally overestimated for the following reason. At the critical point $V=V_c$, $M_2(L)\sim 1/L^{1+\eta}$ for large $L$, where $\eta>0$ is the anomalous dimension of the order parameter. When using $a_0 + \frac{a_1}{L} + \frac{a_2}{L^2}$ to fit $M_2(L)$ at $V=V_c$, negative $a_0$ is generically obtained, which indicates that this fitting could somewhat overestimate $V_c$ when $L$ is not large enough.

\begin{figure}[b]					
\centering
\includegraphics[width=4.25cm]{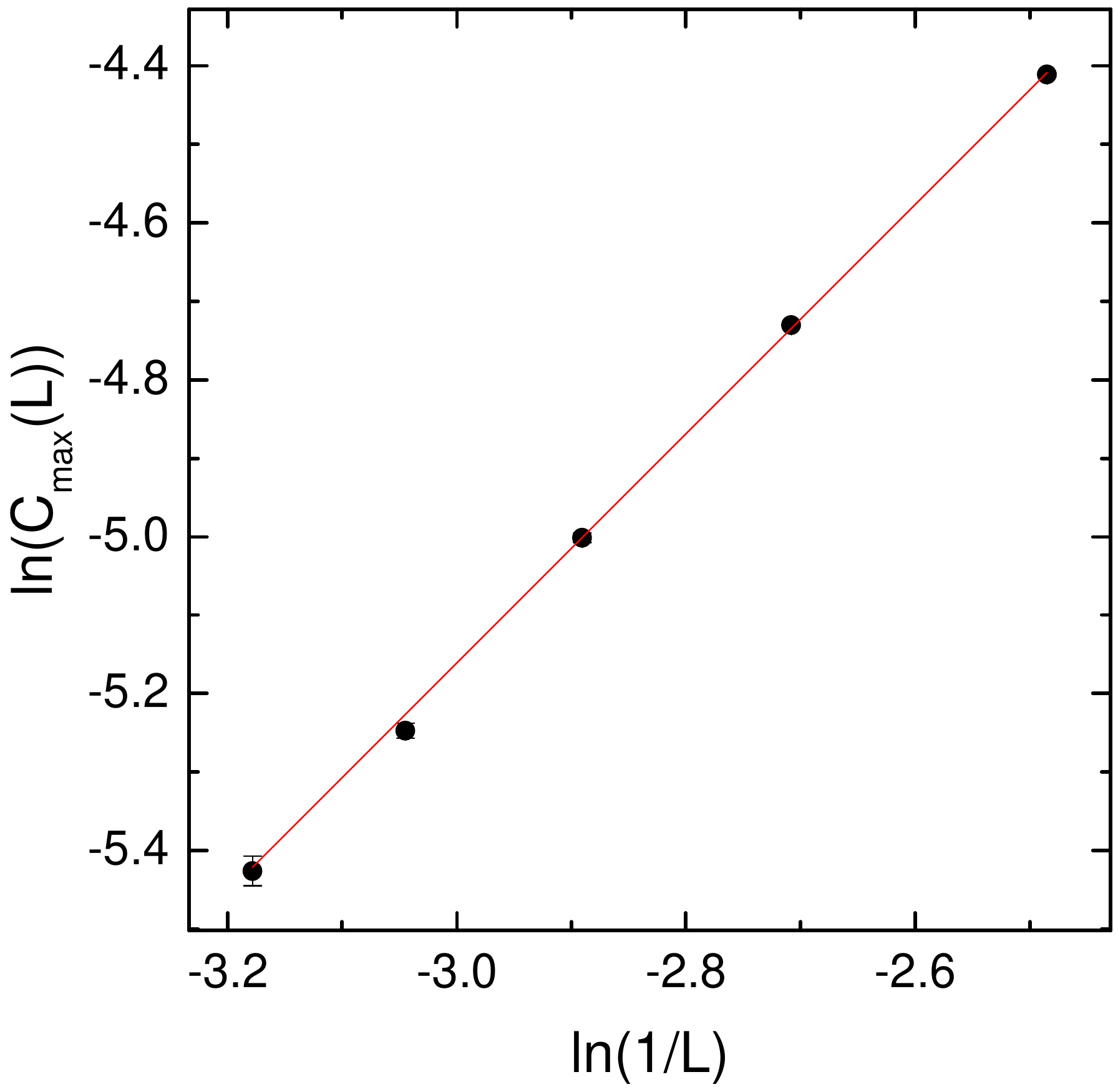}
\includegraphics[width=4.25cm]{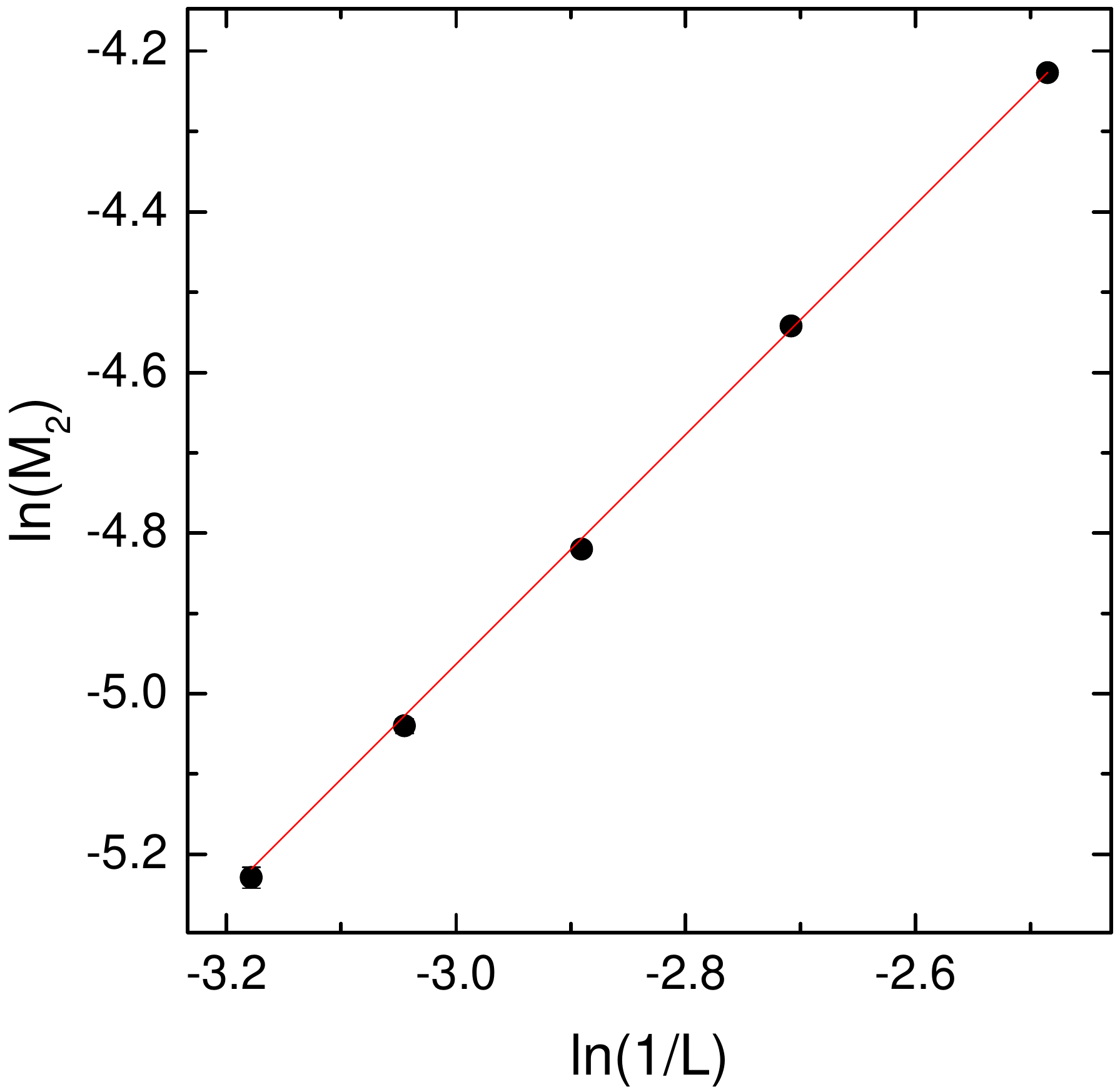}

\caption{(a) For the honeycomb model, the anomalous dimension of the CDW order parameter $\eta=0.45(2)$ is obtained from the finite-size scaling of $C_\textrm{max}(L)$ at $V=V_c$ with $L=12\sim 24$. (b) Similarly, $\eta=0.43(2)$ is obtained from finite-size scaling of $M_2(L)$ of $L=12\sim 24$ at criticality. }
\label{honeycombeta}
\end{figure}

To obtain $V_c$ more accurately, we further compute the quartic of CDW order parameter $M_4$ defined as
\bea
M_4 = \sum_{ijkl}\frac{\eta_i \eta_j \eta_k \eta_l}{N_s^4} \big\langle(n_i - \frac{1}{2})(n_j - \frac{1}{2})(n_k - \frac{1}{2})(n_l - \frac{1}{2})\big\rangle, \nn
\eea
and use the method of Binder ratio $B\equiv \frac{M_4}{M_2^2}$ to determine the critical point of the CDW phase transition. According to the following scaling functions for $M_2$ and $M_4$:
\bea
M_2 &=& L^{-1-\eta} {\cal F}( L^{1/\nu}( V - V_c )), \\
M_4 &=& L^{-2-2\eta} {\cal G}( L^{1/\nu}( V - V_c )),
\eea
where we have implicitly assumed the dynamical critical exponent $z=1$ for the CDW transition in the current model, the Binder ratios of sufficiently large $L$ should cross at $V=V_c$. The calculated Binder ratios for different $V$ and different $L$ are shown in \Fig{honeycombscaling}(b), which clearly show that $V_c\approx 1.355$, which is consistent with the one obtained in Ref.[\onlinecite{LWang-14a}].

\begin{figure}[t]
\centering						
\subfigure{\includegraphics[width=4.25cm]{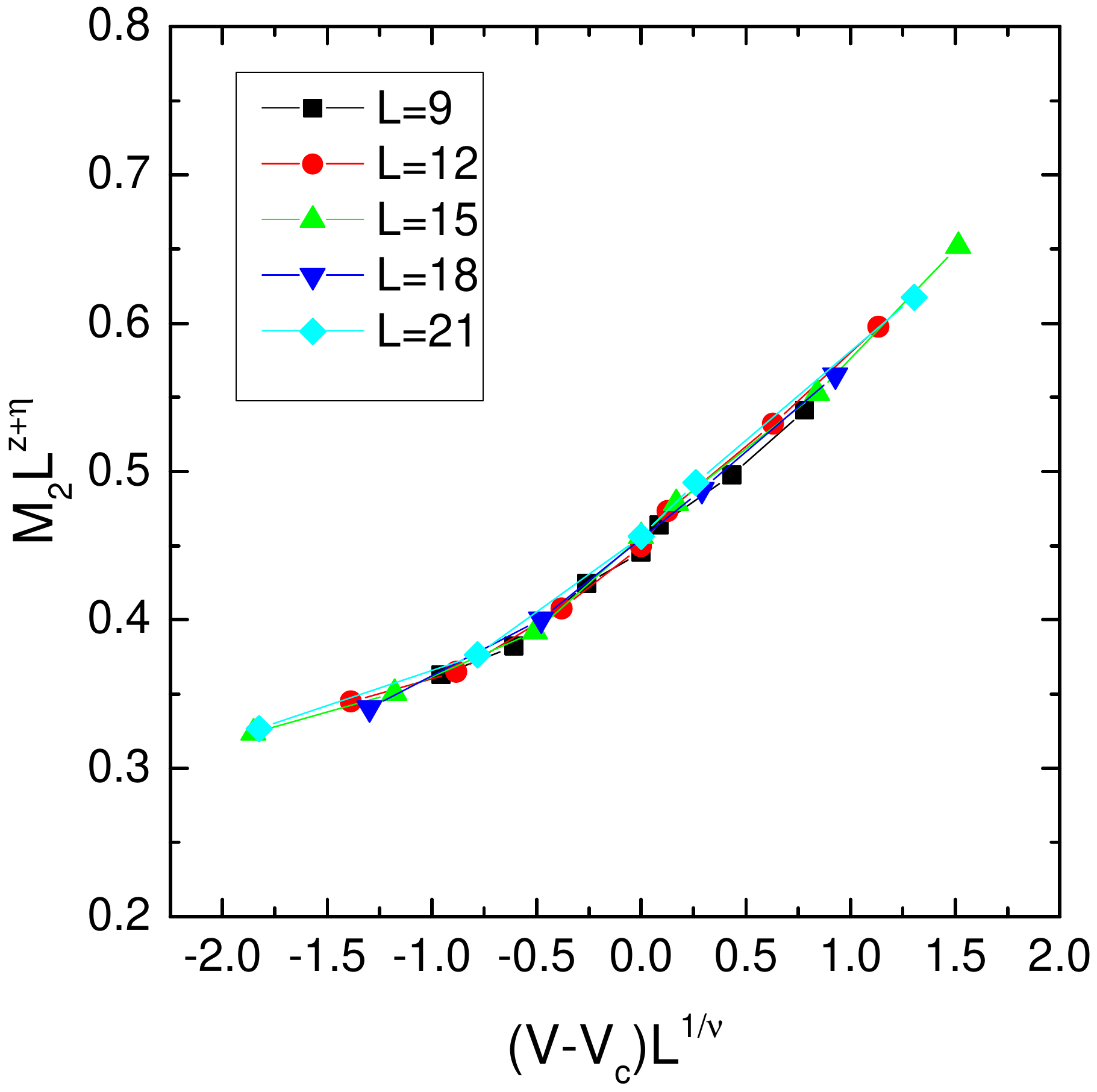}}
\subfigure{\includegraphics[width=4.25cm]{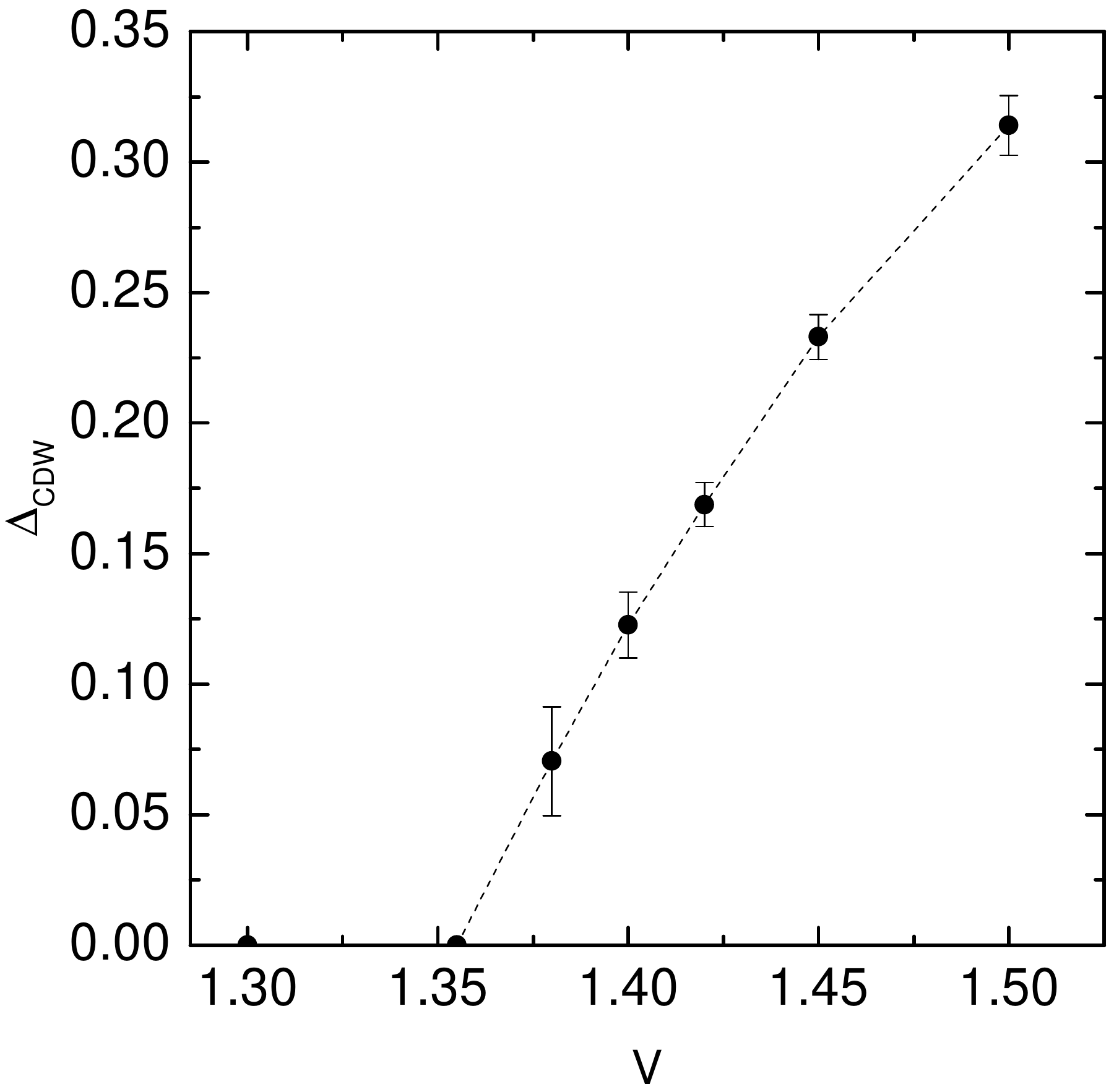}}
\caption{(a) For the honeycomb model, collapsing of data points occurs when $\nu=0.77$. (b) The CDW order parameter at various $V$. We obtain $\beta=0.60(3)$ from fitting the data to $\Delta_\textrm{CDW}\sim (V-V_c)^\beta$. }
\label{honeycombnu}
\end{figure}

After obtaining $V_c$ for the CDW transition, we can further compute the independent critical exponents $\eta $ and $\nu$ which are critical exponents regarding correlation functions. Other critical exponents such as $\beta$ may be obtained from $\eta$ and $\nu$ through hyper-scaling relations. We use two different methods to obtain $\eta$. First, we perform a finite-size scaling analysis of $M_2$ according to the scaling function: $M_2(L) \sim L^{-1-\eta}$ at $V=V_c$. Second, we compute the density-density correlation $C_\textrm{max}(L) = \langle(n_{i}-1/2)(n_{i+\vec r_\textrm{max}}-1/2)\rangle$ where $\vec r_\textrm{max}=(L/2,L/2)$ is the largest possible separation between two sites in the lattice with periodic boundary conditions. At criticality, this correlation decays in power-law as $C_\textrm{max}(L) \sim  1/L^{1+\eta}$ for sufficiently large $L$. In \Fig{honeycombeta}(a) and \Fig{honeycombeta}(b), we plot $M_2(L)$ and $C_\textrm{max}(L)$ versus $1/L$, respectively, in a log-log way, and then fit them by a linear function whose slope is $1+\eta$. From the fitting of $C_\textrm{max}(L)$, we obtain $\eta=0.45(2)$. Slightly smaller but similar $\eta$ within the error bar is obtained from the fitting of $M_2(L)$.

We are ready to obtain the critical exponent $\nu$ using $M_2 L^{1+\eta}={\cal F}( L^{1/\nu}( V - V_c ))$ where ${\cal F}$ is a scaling function. There exists an appropriate $\nu$ such that different points ($M_2 L^{1+\eta}, L^{1/\nu}(V-V_c)$) of various $V$ around $V_c$ and different $L$ should collapse on a single curve ${\cal F}$ even though ${\cal F}$ is an unknown function.  As shown in \Fig{honeycombnu}(a), all data points of different $V$ and $L$ collapse best to a single curve when we choose $\nu=0.77(3)$. From the scaling relation $\beta=\frac{\nu(1+\eta)}{2}$, we obtain $\beta = 0.58$, which is also consistent with the value $0.60$ obtained by fitting the order parameter $\Delta_\textrm{CDW}\sim (V-V_c)^\beta$, as shown in \Fig{honeycombnu}(b).  Consequently, we have shown that $\eta=0.45(2)$ and $\nu=0.77(3)$ for the $N=2$ chiral-Ising universality class in 2+1D, which are reasonably close to $\eta\approx 0.50$ but have slight discrepancy with $\nu\approx 0.88$, obtained in the two-loop RG calculations in the $\epsilon=4-D$ expansion[\onlinecite{Rosenstein-93}].

\subsection{The CDW transition on the $\pi$-flux square lattice}
\begin{figure}[t]						
\centering
\subfigure{\includegraphics[width=4.25cm]{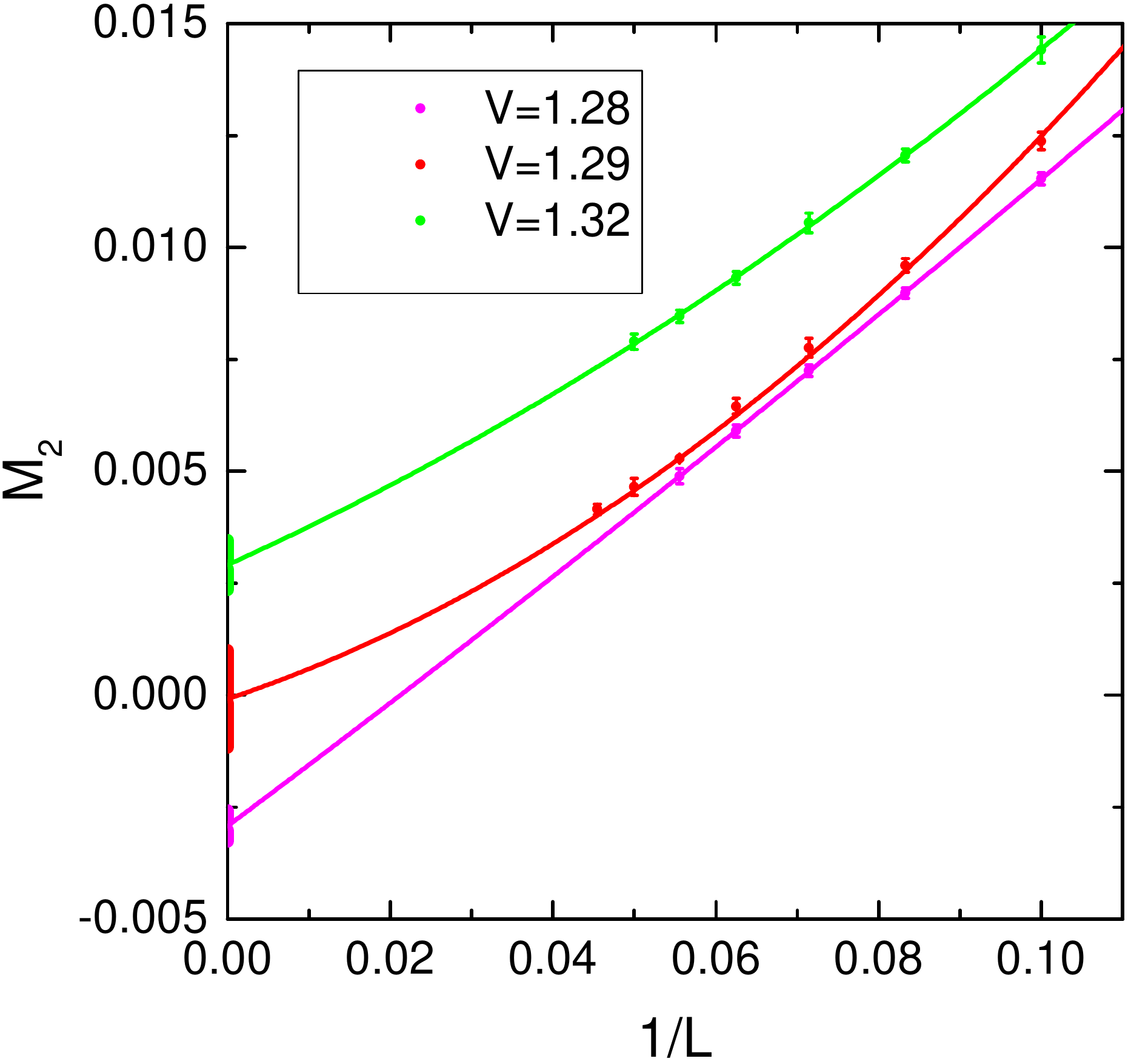}}
\subfigure{\includegraphics[width=4.25cm]{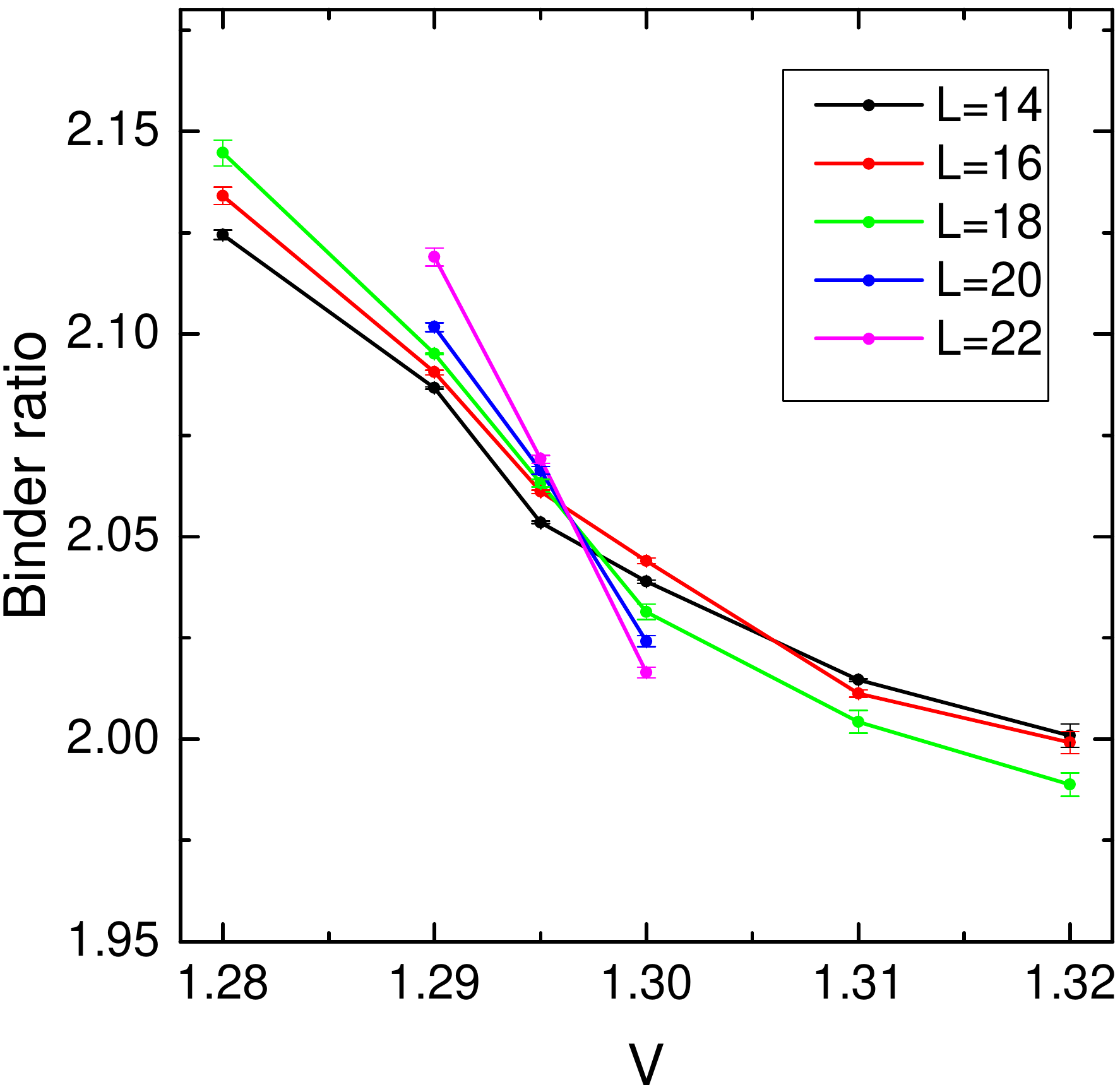}}
\subfigure{\includegraphics[width=4.25cm]{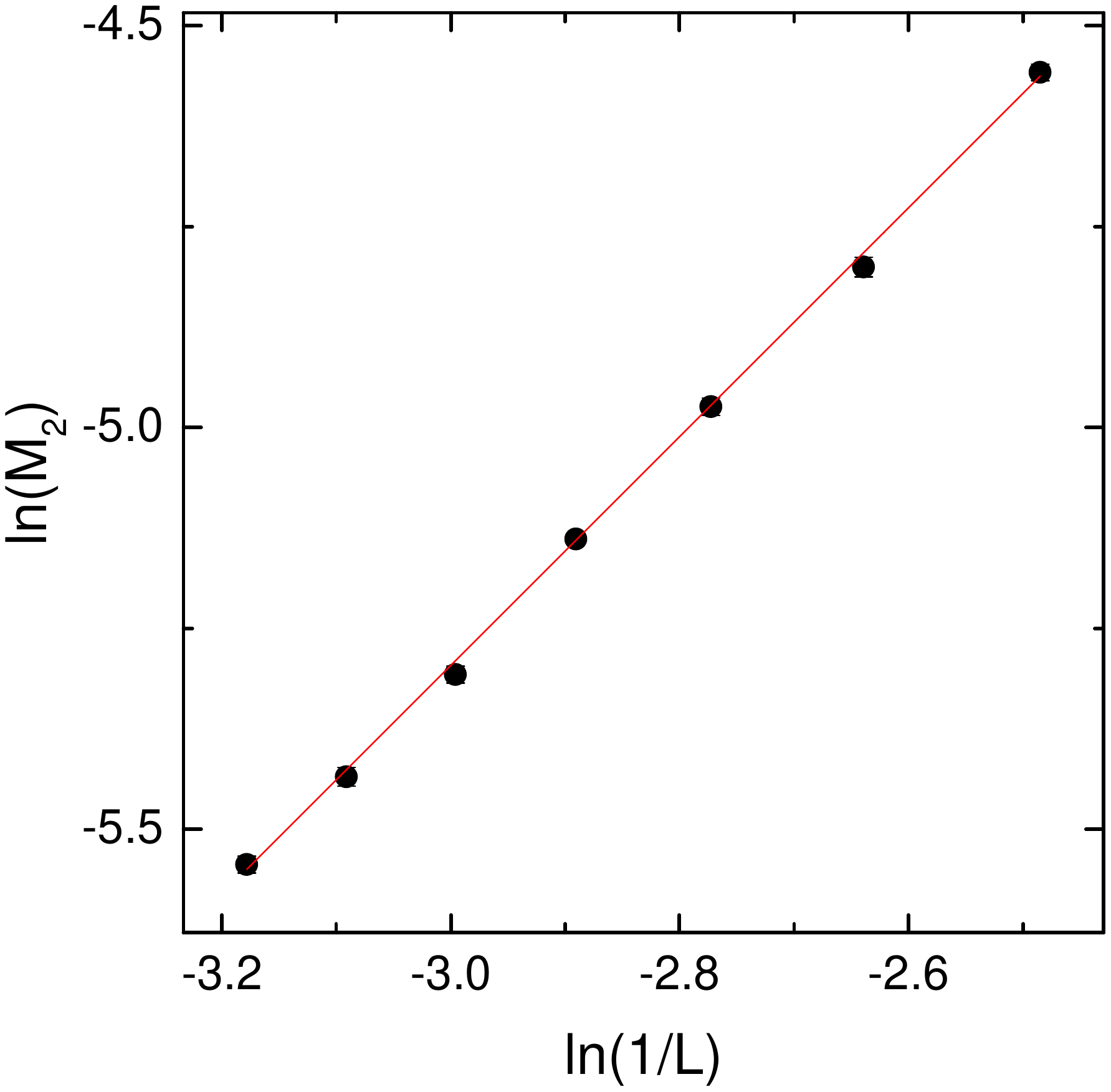}}
\subfigure{\includegraphics[width=4.25cm]{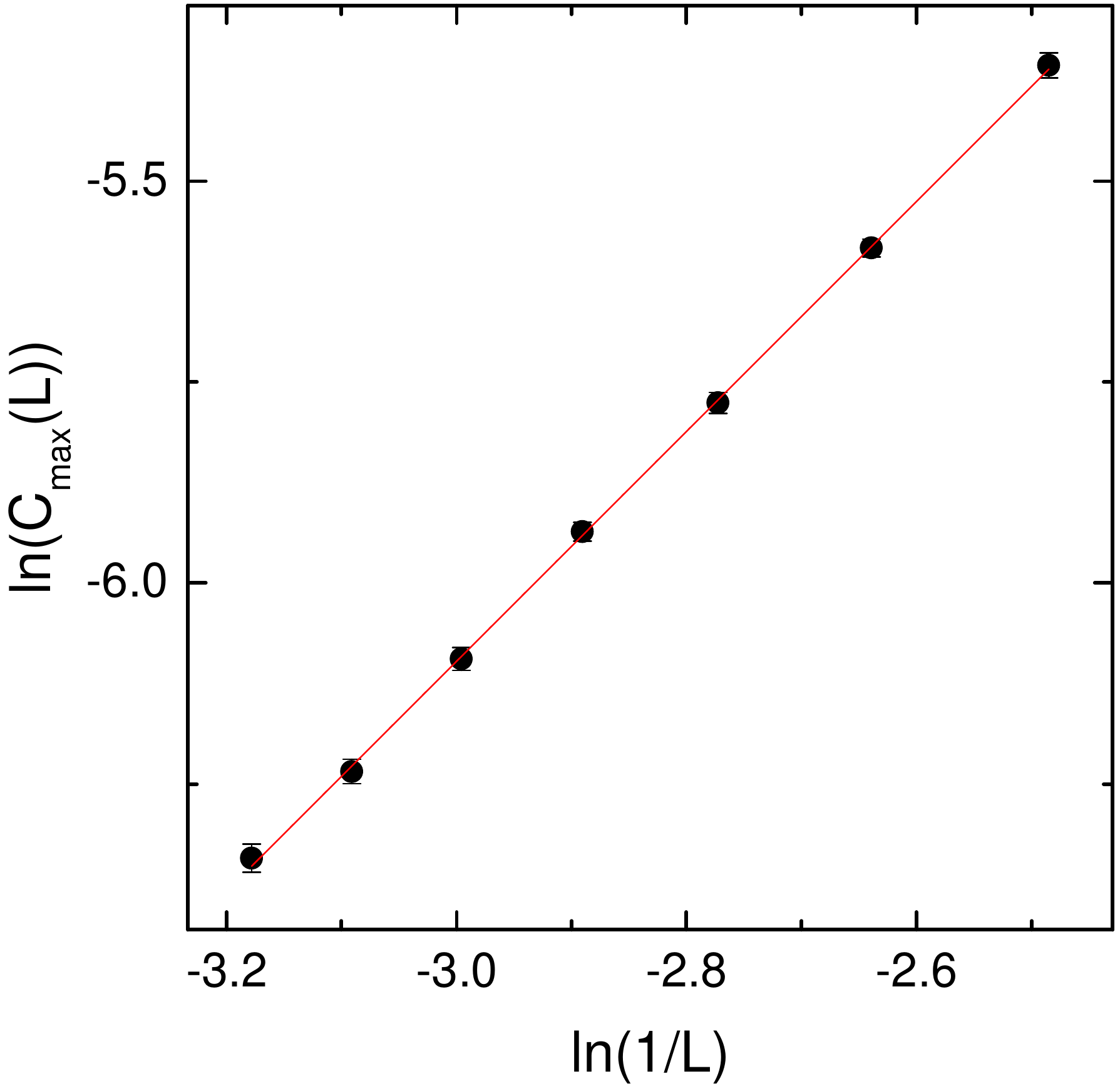}}
\subfigure{\includegraphics[width=4.25cm]{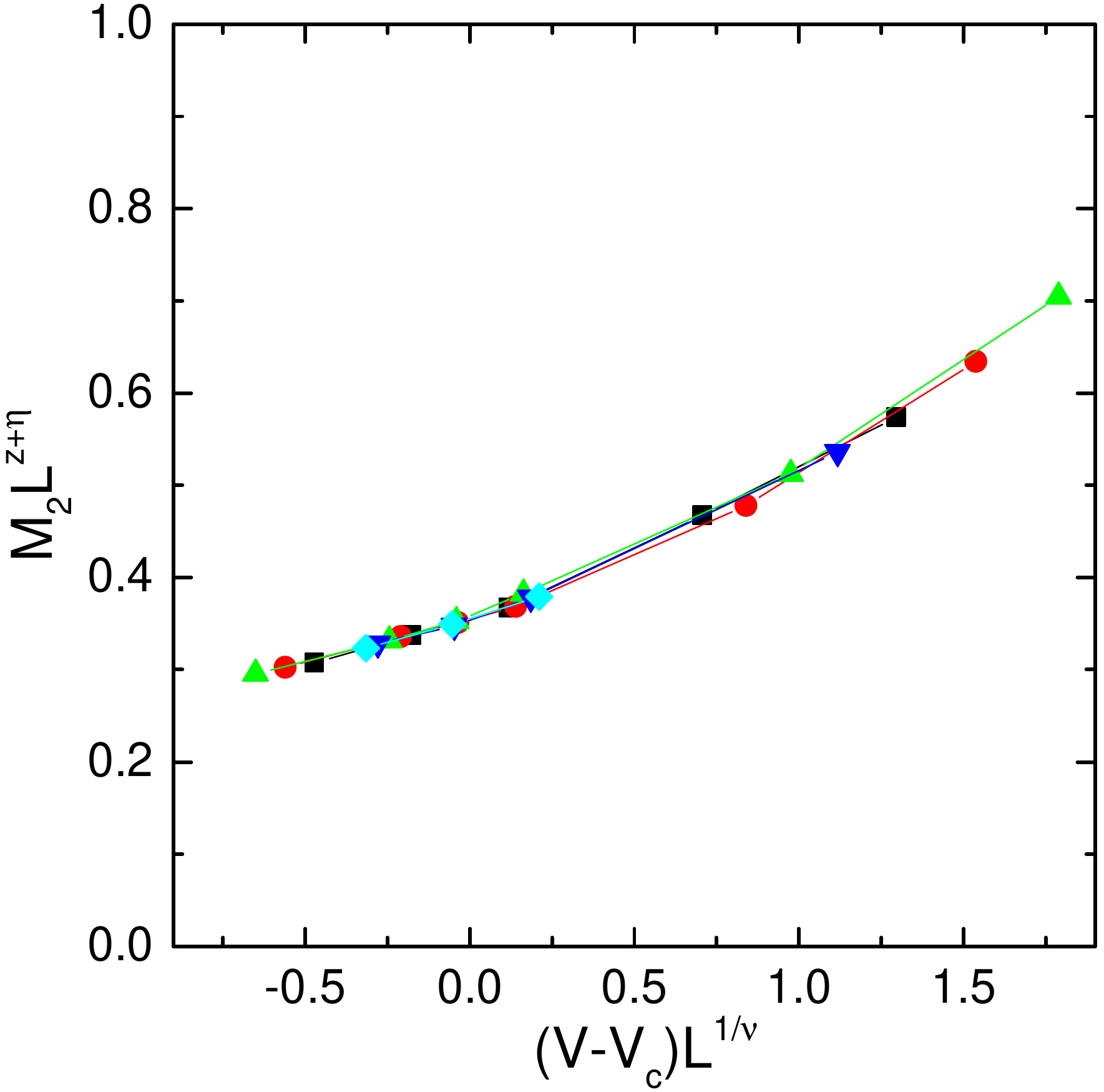}}
\subfigure{\includegraphics[width=4.25cm]{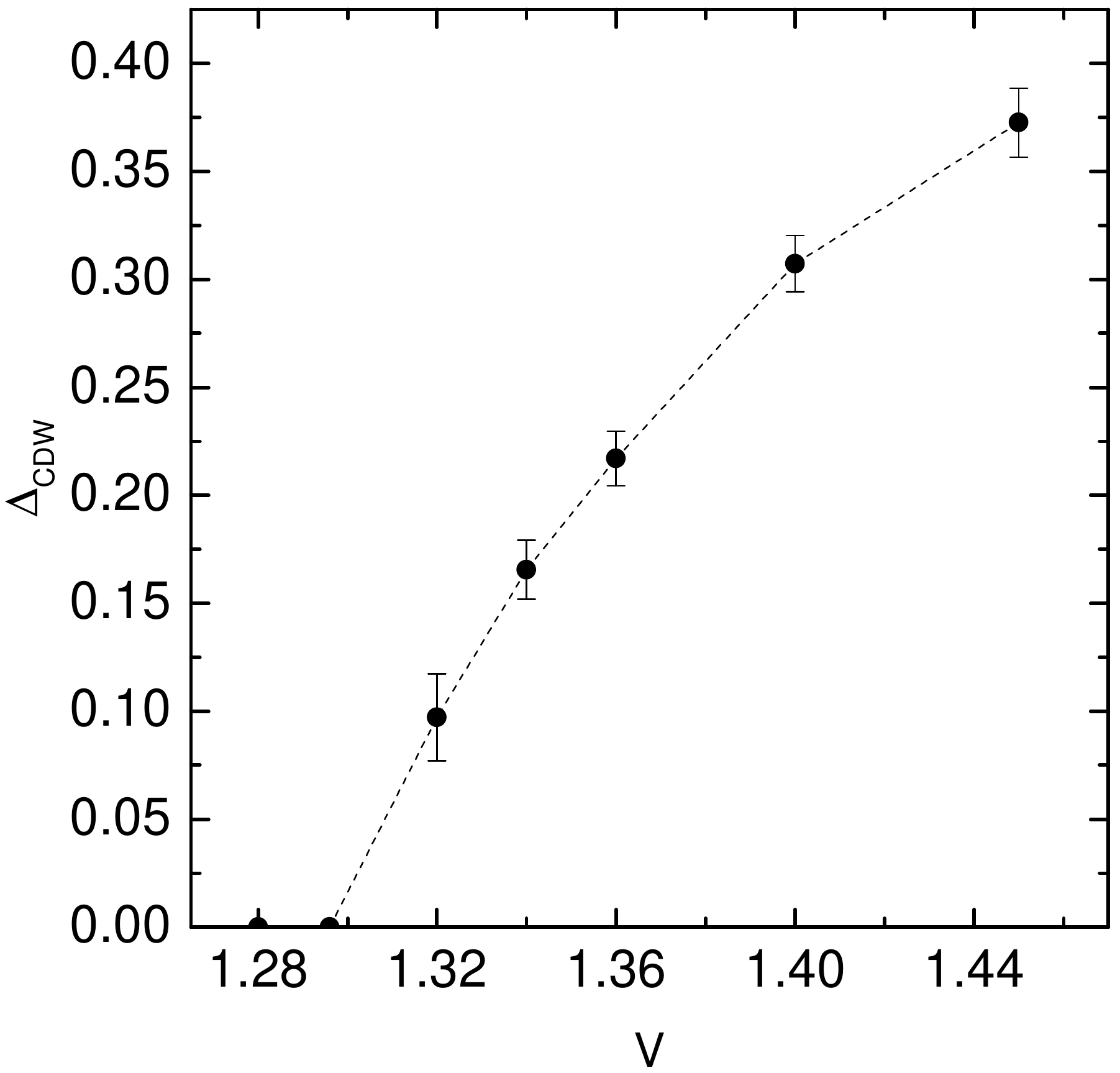}}
\caption{MQMC simulations of the $\pi$-flux square model. (a) Finite-size scaling of the CDW structure factor $M_2$ obtained in projector (zero-temperature) MQMC simulations on lattice of $N_s=2L^2$ sites for various $V$ and $L=6\sim 21$; (b) The Binder ratios $B\equiv M_4/M_2^2$ for various $V$ and $L=9\sim 21$. The crossing of Binder ratios shows that the critical interaction for the CDW transition is $V_{c}=1.296(1)$; (c) The anomalous dimension of the CDW order parameter $\eta=0.43(2)$ is obtained from the finite-size scaling of $C_\textrm{max}(L)$ at criticality. (d) Similarly, $\eta=0.42(2)$ is obtained from finite-size scaling of $(M_2(L))$ at criticality. They are consistent within error bar. (e) Collapsing of data points occurs when $\nu=0.79$; (f) The CDW order parameter at various $V$. We obtain $\beta=0.67(4)$ from fitting the data to $\Delta_\textrm{CDW}\sim (V-V_c)^\beta$.  }
\label{squarescaling}
\end{figure}

It has been known that fermions on the square lattice with $\pi$-flux per plaquette also feature dispersions of massless Dirac fermions. Spinless fermions on the square lattice with $\pi$-flux per plaquette and with NN repulsive interactions are described by the same Hamiltonian as \Eq{ham1} in which $t_{ij}=e^{i\theta_{ij}}t$ with $\sum_{\avg{ij}\in \square}\theta_{ij}=\pi$ (mod $2\pi$). At zero temperature, a similar CDW transition occurs when the NN interaction $V$ exceeds a critical value $V_c$ and this CDW transition is in the same universality class of the $N=2$ chiral-Ising transition in 2+1D. To investigate the quantum critical behavior of this CDW transition, we performed fermion-sign-free MQMC simulations of this model, did similar data analysis, and expected to obtain identical critical behaviors. The results are shown in \Fig{squarescaling}. From the Binder ratio analysis, we obtain $V_c=1.296(1)$. Moreover, finite-size scaling renders $\eta=0.43(2)$, $\nu=0.79(4)$, and $\beta=0.67(4)$, which are consistent with the corresponding values obtained for the honeycomb model, within error bar.

\section{Conclusions and remarks}
Using the fermion-sign-free MQMC method, we have simulated the spinless fermion model on the honeycomb lattice as well as on the $\pi$-flux square lattice with NN repulsions. We focused on the quantum critical behavior at the CDW transition in these models, which is in the universality class of the chiral-Ising transition of $N=2$ two-component mass Dirac fermion in 2+1D. The low energy physics of this CDW transition can be described by the $N=2$ Gross-Neveu or Gross-Neveu-Yukawa theory in 2+1D. We numerically showed that $\eta=0.45(2)$, $\nu=0.77(3)$, and $\beta=0.60(3)$ at this $N=2$ chiral-Ising transition in 2+1D.

The critical exponents obtained by the fermion-sign-free MQMC simulations are reasonably consistent with the ones obtained by the two-loop RG calculations in $\epsilon=4-D$ expansion even though there is still some slight discrepancy. The discrepancy should not come from the differences in the models used in numerical simulations and in RG calculations. In the RG calculations, the two massless Dirac fermions have the same chirality, which means that the symmetry breaking phase in its corresponding lattice model is the quantum anomalous Hall state. The two Dirac fermions on the honeycomb lattice have opposite chirality and the broken symmetry phase is a CDW state. Nonetheless, the RG equations of the coupling constants in the two corresponding low-energy field theories describing the QAH and CDW transitions are identical. Consequently, the two seemly-different transitions should be in the same universality class, at least, in the sense of the critical exponents even though the two broken symmetry phases are topologically distinct. We believe that the critical exponents obtained in the present MQMC simulations could serve as a benchmark for more accurate higher-loop RG calculations in the $\epsilon$-expansions which are deferred to future studies.

{\bf Acknowledgement:} We would like to thank Fahker Assaad, Igor Herbut, Ziyang Meng, Matthias Troyer, and Lei Wang for helpful discussions as well as the National Supercomputer Center in Guangzhou of China (Milky Way II supercomputer system) for computational support. This work is supported in part the the National Thousand-Young-Talents Program (H.Y.) and by the NSFC under Grant No. 11474175 (Z.X.L, Y.F.J., and H.Y.).

\end{document}